# Damage identification on spatial Timoshenko arches by means of genetic algorithms

A. Greco[a], D.D'Urso[a], F. Cannizzaro[a], A. Pluchino[b]

[a] Department of Civil Engineering and Architecture, University of Catania, viale A. Doria 6, Catania, Italy

[b] Department of Physics and Astronomy, University of Catania, and INFN Sezione di Catania, viale A. Doria 6, Catania, Italy

**Abstract**

In this paper a procedure for the dynamic identification of damage in spatial Timoshenko arches is presented. The proposed approach is based on the calculation of an arbitrary number of exact eigen-properties of a damaged spatial arch by means of the Wittrick and Williams algorithm. The proposed damage model considers a reduction of the volume in a part of the arch, and is therefore suitable, differently than what is commonly proposed in the main part of the dedicated literature, not only for concentrated cracks but also for diffused damaged zones which may involve a loss of mass. Different damage scenarios can be taken into account with variable location, intensity and extension of the damage as well as number of damaged segments. An optimization procedure, aiming at identifying which damage configuration minimizes the difference between its eigen-properties and a set of measured modal quantities for the structure, is implemented making use of genetic algorithms. In this context, an initial random population of chromosomes, representing different damage distributions along the arch, is forced to evolve towards the fittest solution. Several applications with different, single or multiple, damaged zones and boundary conditions confirm the validity and the applicability of the proposed procedure even in presence of instrumental errors on the measured data.

# 1 Introduction

The presence of damage in beam-like structures, either straight or curve, implies a loss of the structural stiffness inducing variation of both static and dynamic responses. Its early identification allows preventing the occurrence of more severe damage or even structural failures and has therefore been the object of several studies in the last decades. The effects of damage and its identification have been investigated in the literature, mainly with reference to straight beams, using different techniques. These strategies are mainly based on the variation of either dynamic characteristics, such as natural frequencies [1-10], mode shapes [11-13], or static quantities, such as displacement induced by applied loads [14-17].

The most interesting application of these techniques can be found in the health monitoring applications, whereby the use of experimental data from non-destructive tests allows identifying damage [18-20].

Damage can affect large or small areas; according to the involved damaged area and to the depth of the damage several models are available in the literature [21-30]. When the damage is located in a restricted area, the most used mathematical model to simulate its presence is the so called equivalent hinge model, whose calibration can be performed according to numerous proposals [28,31]. The adoption of this model can be performed either by enforcing continuity conditions at the cracked sections, or by means of suitable closed form solutions depending on the two ends boundary conditions only; such an approach has been recently extended to the statics of multi-cracked circular arches [32]. Such models, although appropriate for concentrated damage, are not adequate in case the damaged zone has a certain extent and involves a loss of mass. Recently, two of the authors of this manuscript, considering the same strategy here adopted to model the damage, investigated the direct problem of the dynamics of damaged arches, bringing to light some interesting properties related to the mass variation in the damaged zone [33].

Many papers on the dynamic damage identification in curved beams have been published in the scientific literature, the largest part of these publications deal with circular arches [34,35] nevertheless some studies have been presented concerning the vibration of arches with variable curvature [18,19,36]. A recent overview on damage identification techniques via modal curvature analysis has been given by Dessi and Camerlengo [37].

A recent area of study in damage identification deals with genetic algorithms which are inspired by Darwin's theory and are based on the process of natural selection. These algorithms are able to provide robust tools for solving optimization problems and explore the region of interest by running several times the same procedure with different initial conditions in order to locate with high probability the global optimum avoiding local minima. Many papers have been published dealing with the use of genetic algorithms in damage identification on straight beams [38-43].

In the present paper, spatial arch structures, in which the damage is modelled as a reduction of the cross section in a concentrated zone [33,44], are studied. Differently than the widely adopted damage model with equivalent concentrated hinges and rotational springs, the proposed procedure is able to identify also finite portions of damaged arch and also to consider the possible reduction of the mass of the structural element. The inverse problem of identifying damage parameters is solved by means of genetic algorithms implementing an optimization procedure based on the minimization of an objective function which measures the difference between calculated and experimental values of an assumed number of natural frequencies and modes of vibration. To the authors' best knowledge this is the first application of genetic algorithms in damage identification with reference to spatial arches.

The exact natural frequencies of vibration and the corresponding modes are calculated by means of an application of the Wittrick and Williams algorithm [45] through a procedure introduced by Howson et al. [46] and revisited by the authors in a previous paper [47] in which the dynamics of undamaged spatial arch structures has been exactly investigated in the context of the dynamic stiffness method [48].

Multiple damage portions of the arch are here taken into account, thus demonstrating the suitability of the model also in presence of complex damage scenarios. The proposed procedure is able to identify four parameters for each damage, namely its location, its intensities along the two main directions and its extension. The robustness of the procedure has been tested in presence of noise and through a comparison with an experimental test available in the relevant literature. In addition, considerations on the number of data required for an accurate damage identification, and on the use of the mode shapes combined with the natural frequencies, have been formulated.

## 2  The eigen-properties of spatial arches

In the free vibration of curved beams either in-plane or out-of-plane motions may occur, where the former consists primarily of bending-extensional modes while the latter is essentially bending-twisting dynamics [47]. In the present section the dynamics of spatial Timoshenko arches, either in-plane or out-of-plane, is briefly summarized.

Let an infinitesimal Timoshenko arch element, subtending an angle $d\varphi$ at the centre of a circle of radius $R$, be considered as shown in Figure 1. The cross-section is assumed to be symmetric with respect to the two directions denoted as 1-1 and 2-2, and its properties are: area $A$, second moments of area $I_1$ and $I_2$, shear correction factor $k'$, torsional constant $J$ and polar moment of inertia $I_p$, all of them assumed to be constant. Furthermore the material properties, Young's modulus $E$, shear modulus $G$ and density per unit length $\rho$, are uniformly distributed.

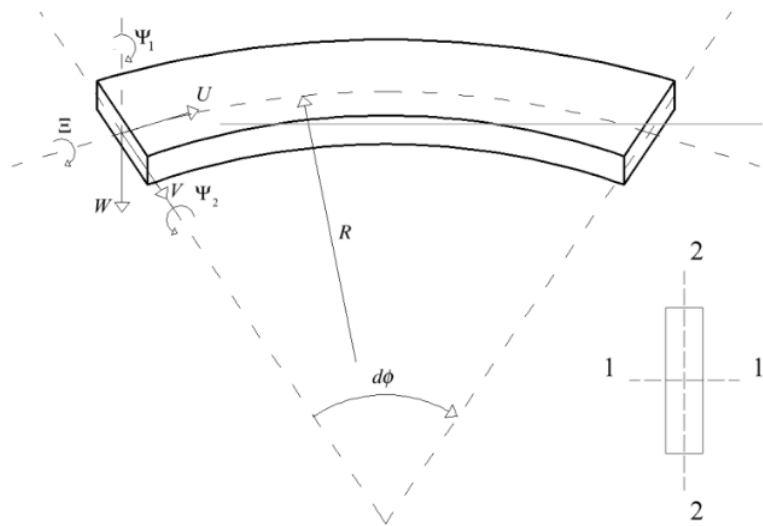

*Figure 1 - The infinitesimal arch element and its cross section*

The degrees of freedom, in the space, with reference to both the in-plane and the out-of-plane motions, of the generic cross section are reported in Figure 1. By considering the equilibrium of the arch element, in its undeformed configuration, and the linear elastic constitutive law, the differential equations of motion governing the in-plane free vibration of the Timoshenko arch in a useful dimensionless form, have been presented in a recent paper [47] and are here briefly summarized. The assumption that, in the undeformed configuration, the arch is contained in a principal plane of the cross section guarantees that in-plane and out-of-plane vibrations are governed by uncoupled differential equations [46].

With reference to the kinematic parameters shown in Figure 1, the equations governing the in-plane motion of the arch are:

$$D(D^2 + 1 + \gamma^2)U - (D^2 + 1 - \gamma^2)V = 0$$

$$\{\bar{v}D^4 - (1 + \lambda^2 - 2\bar{v}\gamma^2)D^2 - \gamma^2(\lambda^2 + 1 - \bar{v}\gamma^2)\}U +$$
$$-\{D[(1+\bar{v})D^2 - \lambda^2 + \gamma^2(1+\bar{v})]\}V = 0 \tag{1}$$

where the symbol $D^n$ denotes the $n$-th derivation with respect to the polar coordinate $\phi$, and the following dimensionless parameters have been introduced:

$$\lambda_1^2 = \frac{AR^2}{I_1} \qquad \gamma^2 = \frac{\rho\omega^2 R^2}{E} \qquad \bar{v} = \frac{E}{k'G} \tag{2}$$

$\omega$ being the frequency of vibration. The differential equations governing the dynamic out of plane equilibrium of the considered Timoshenko arch turn out to be:

$$\left(D^4 + \left[(1+\bar{v})\gamma^2 - \mu\right]D^2 - \left(\lambda_2^2 + \bar{v}\mu - \bar{v}\gamma^2\right)\lambda_2^2 \gamma^2\right)W +$$
$$+\left(-D^2(1+\mu)\right)\Xi = 0$$
$$\left(D^2(1+\mu) + \bar{v}\gamma^2(1+\mu)\right)W - \left(-\mu D^2 + 1 - \eta\gamma^2\right)\Xi = 0 \tag{3}$$

where:

$$\mu = \frac{GJ}{EI_2} \qquad \eta = \frac{I_p}{I_2} \qquad \lambda_2^2 = \frac{AR^2}{I_2} \tag{4}$$

The knowledge of the solution of the equations of motion described in [47] allows the evaluation of the exact dynamic stiffness matrix of the arch which relates the vector **P** of the nodal static parameters at the end of the element, to the corresponding vector of kinematic parameters **d** through

the equation:

$$\mathbf{P} = \mathbf{K}(\omega)\mathbf{d} \tag{5}$$

where $\mathbf{K}(\omega)$ can be obtained by assemblage of the dynamic stiffness matrices of all the members of the structure. Once the dynamic stiffness matrix of the structure is assembled, the natural frequencies can be calculated by means of a very effective method based on the Wittrick & Williams Algorithm [45]. The algorithm allows to evaluate the number of frequencies of vibration which are lower than a trial value and, therefore, by means of an iterative procedure, to converge to any required frequency.

The correspondent $i$-th mode of vibration ($i$=1, 2, 3…) can be evaluated calculating for each kinematic parameter $X_i$ the following expression:

$$X_i(\phi) = \left( \sum_{m=1}^{6} A_m e^{\beta_m \phi} \right)_i \tag{6}$$

where the constants $A_m$ and $\beta_m$ for the generic $i$-th mode are calculated for in-plane and out of plane behaviours when the boundary conditions for the single arch element are assigned [47].

## 3 The model of the damaged arch

In order to perform an accurate modelling of a structural damage many parameters should be taken into account. In fact two damaged parts of a structural element may differ for dimension, length, shape and position of the damaged zone. Many studies in the literature consider that the damage can be assumed to be concentrated and can be modelled by means of a rotational spring of suitable rigidity.

In the present paper the damage is simply modelled introducing an element with a cross section of the same shape of the undamaged one but a smaller size. This element has a weaker stiffness and a smaller mass with respect to the correspondent intact element and is therefore able to model a reduction of the mass in the damaged portions with respect to the healthy condition [33].

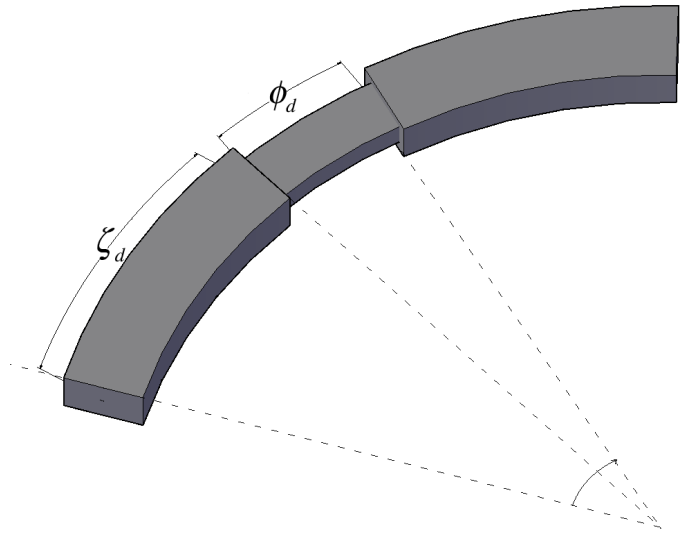

*Figure 2 - The damaged element*

In general, each damage can be described according to four damage parameters, as highlighted in Figure 2:

- $\zeta_d$: position of the damaged element, defined as the curvilinear abscissa of the left reduced cross section normalized with respect to the total arch length;

- $\phi_d$: extension of the damaged zone, defined as a ratio of the opening angle;

- $\beta_l$ ($l$=1, 2): cross section intensity of the damage, defined as the ratio between the dimension of the undamaged and damaged cross section either in the 1-1 or in the 2-2 directions.

It is worth to notice that in general two independent stiffness reductions corresponding to the two principal axes of the section should be considered, in the latter case each damaged part would need two intensity parameters to completely define the damage implying a total amount of four parameters associated to each damage. In the following a homothetic reduction of the cross section along the two principal directions will be considered, thus requiring only one parameter $\beta = \beta_1 = \beta_2$ to completely identify each damaged part in terms of stiffness reduction.

When the extension of the damage is close to zero, no crack closure phenomenon is considered, hence linear behaviour of the damaged curved beam is always assumed.

# 4 Procedure for the damage identification

In the proposed damage identification procedure, the eigen-properties of spatial arches, calculated as described in the previous paragraphs, are taken into account. The generic damage scenario is associated to several damaged zones, each one characterized by its location, its extension and its intensity. Therefore in the inverse problem three different parameters have to be identified for each damaged segment of the arch.

In order to seek for the optimal damage configuration, an arbitrary number $N$ of calculated natural frequencies $\omega_j^{calc}$ and modal shapes $\varphi_j^{calc}$ ($j=1,..N$) can be computed and compared with the correspondent measured values from experimental tests $\omega_j^{exp}$, $\varphi_j^{exp}$ in order to minimize the differences. The mode shapes are computed at $N_m$ given polar coordinates $\phi_k$ which represent the measurement points used for the data acquisition. Since several damage identification procedures are frequency based only, it is interesting to investigate on the additional contribution of the modal shapes. Therefore, in the following two different objective functions will be considered in the optimization, namely a function which considers the contribution of the natural frequencies only ($O_1$), and a function which accounts for both frequencies and modal shapes ($O_2$) according to the Modal Assurance Criterion (MAC) [38,49]. The objective function $O_2$ has the following expression:

$$O_2(\omega_j,\varphi_j) = W_f \sqrt{\sum_{j=1}^{N}\left[\frac{\omega_j^{calc}-\omega_j^{exp}}{\omega_j^{und}}\right]^2} + W_m \sum_{j=1}^{N} \frac{\left|\varphi_j^{calc} \cdot \varphi_j^{exp}\right|^2}{\left(\varphi_j^{calc} \cdot \varphi_j^{calc}\right)\left(\varphi_j^{exp} \cdot \varphi_j^{exp}\right)} \qquad (7)$$

where $\omega_j^{und}$ ($j=1,..N$) are the natural frequencies of the undamaged arch $W_f$ and $W_m$ are appropriate weighting factors. The objective function $O_1$ can be recovered from Eq. (7) when $W_m=0$. It is known that mode shapes are less sensitive to damage severity than frequency variations, thus they are not a good damage indicator when differences between undamaged and damaged configurations are small. Nevertheless, the addition of data provided by mode shapes, although difficult to measure, might be of great help in the damage identification since the frequencies are global scalar parameters, whereas the mode shapes measured at a certain numbers of sections add a quite larger amount of data to be employed in the procedure. In addition, whenever eigen-modes can be measured, they are able to avoid any symmetry in the identified damaged configurations.

Once experimental measures are at disposal, the minimization problem provided in Eq.(7) can be solved by seeking for the optimal damage scenario which best approximates the actual dynamic response of the damaged arch. In this paper the inverse problem of identifying the damage parameters is approached by discretizing the axis of the arch into an arbitrary number $N_p$ of parts. The length of this generic segment also defines the minimum extension of the damage that the procedure will be able to detect. Each part can be considered damaged with different cross sections intensities of the damage $\beta$ previously introduced, as shown in Figure 3.

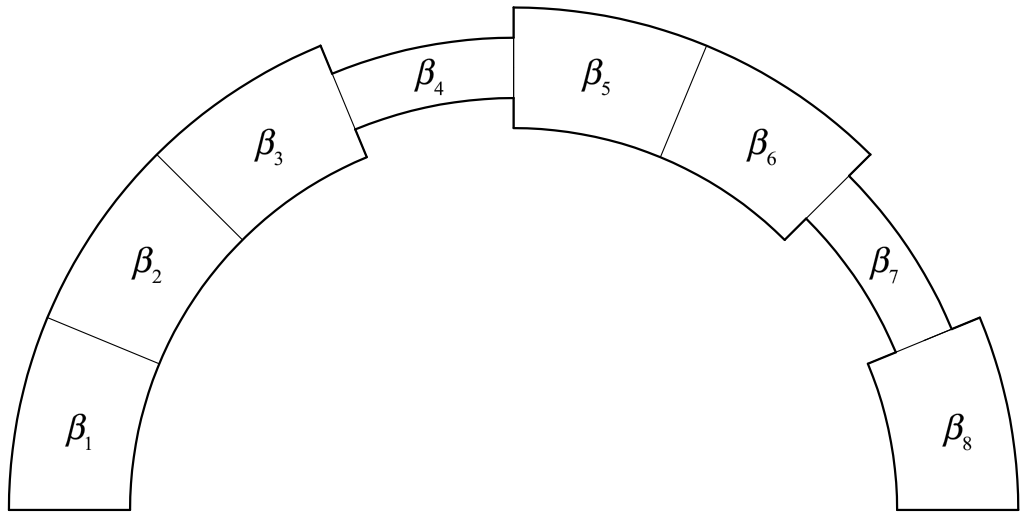

*Figure 3 - The considered arch model for damage identification*

The location of each damaged zone corresponds with the position of its first node (left), its magnitude is characterized by the $\beta$ parameter and the extension is associated to the number of contiguous damaged parts with the same value of $\beta$. In particular, when no damage is associated with a part of the arch, the correspondent value of $\beta$ is equal to 1. Furthermore, $\beta$ assumes values greater than one, proportional to the intensity of the damage, for the damaged parts of the arch.

In order to reduce the computational effort related to the application of the Wittrick and Williams algorithm, it is convenient to minimize the number of parts of the model $N_p$ assembling together contiguous elements with the same value of *β* (Figure 4).

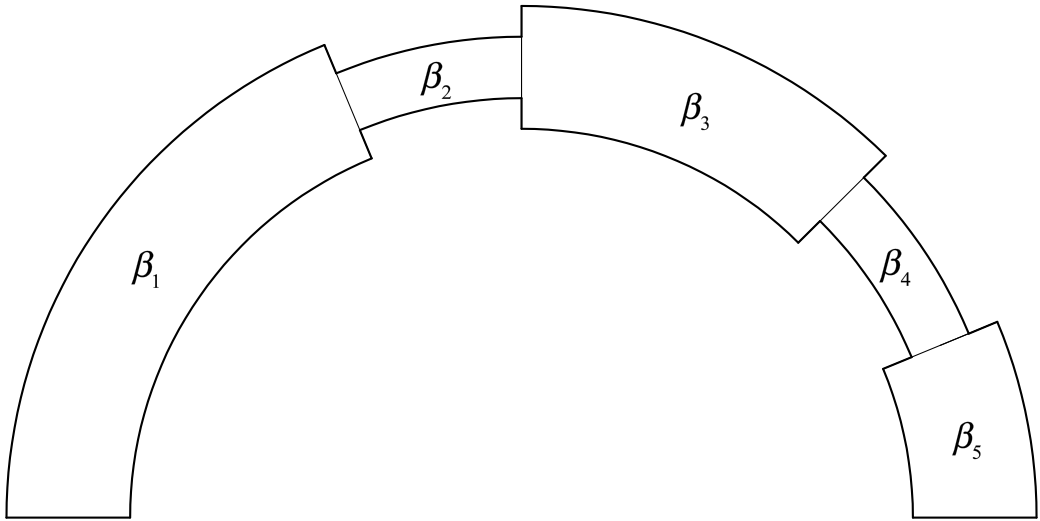

*Figure 4 - The arch model with a reduced number of parts $N_p=5$*

It is clear that the size of the problem grows exponentially with the number of parts of the arch since the dynamic stiffness matrix of the model has size $6(N_p+1)$ and therefore the algorithm is faster when the arch is less damaged.

Since the total number of possible damage configurations, with different locations and intensity, is very large, particularly when the arch is discretized into a great number of curved elements, it is extremely important to have a procedure able to provide the optimal solution within reasonable computing time. For this reason, an optimization method is required and, as better explained in the next section, a genetic algorithm strategy is adopted to select the damage scenario, which best fits with the experimental data.

## 5 Optimization procedure by means of Genetic Algorithms

In this paragraph the optimization procedure adopted in the present paper, which makes use of genetic algorithms, is described.

A "genetic algorithm" is an adaptive stochastic method that mimics the Darwinian evolution, based on an opportune combination of random mutations and natural selection, in order to find optimal numerical values of some specific functions. The algorithm acts over a population of *P* potential solutions (initially randomly chosen) by applying, iteratively, the "survival of the fittest" principle: in such a way it produces a sequence of new generations of individuals that evolves towards a stationary population where the large majority of surviving solutions do coincide and approach as much as possible the real solution of a practical problem. To this regard this evolutionary strategy has been widely adopted for dynamic identification purposes of straight beams [39-42].

In order to translate into this scenario, the original problem of finding the damage parameters in the considered arch, the curvilinear abscissa is discretized into an arbitrary number of parts $N$, each one individuated by the abscissa of its first node $\zeta_k$ ($k=1…N$). Each individual of the population, called chromosome, is coded as a string of numbers $\beta_k$ ($k=1…N$), called genes, each one representing the intensity of the damage present in the $k$-th part of the arch. The value $\beta_k=1$ indicates no reduction of the original cross section.

A generic chromosome $C_i$ of the population ($i = 1, …, P$) can therefore be coded, according to the scheme reported in Figure 5, in the following string:

$$C_i \equiv (\beta_1, \beta_2, \beta_3, …, \beta_k, …, \beta_N)$$

The cross section intensity of the damage for each gene $\beta_k$ can assume $M$ arbitrary discrete values, from 1 to a maximum intensity $\beta_{max}$, $\beta_k \in [1,…,\beta_{max}]$. The number of genes related to damaged zones, those with $\beta_k > 1$, can be evaluated for each chromosome and will be denoted as $\Sigma$. From combinatory algebra it follows that the overall number of possible different chromosomes is $P_{max} = (M + 1)^N$, a quantity which rapidly increases with $N$ even for small values of $M$. It is evident that, for each chromosome, it is possible to calculate the value of the objective functions reported in Eq. (7) by assuming the presence of $\Sigma$ damages in the string $C_i$ with positions given by the coordinates of the corresponding parts and intensities given by the values of the corresponding genes. With this approach it is also possible to consider in the damaged scenario the extension of the damage. In fact the length of the damaged zone is the sum of contiguous parts with the same value of $\beta_k$.

The codification of the damage scenario here proposed is deeply different from similar studies, concerning straight beams, proposed in the literature.

In fact, in these studies the number of notches has to be known *a priori* since each chromosome has a fixed length 2Σ. The latter aspect is due to the structure of the chromosome, which contains, for each damage assumed to be present in the structural element, its position and intensity. Adopting the proposed procedure instead (based on a segmental mesh of the element where each gene represents the level of integrity of the corresponding segment) the number of possible identifiable damaged areas is theoretically equal to the segments of the considered mesh. As a consequence, the number of possible damaged areas to be identified is limited by the amount of available data only; in fact, for each damaged area three or four parameters have to be identified, that is position extension and intensity (a single intensity in case of homothetic reduction of the section, two in case of independent reductions in the two principal directions of the section). Therefore, three eigen-frequencies (or four in case of independent section reductions) are needed for each damage to be identified. As an example, if nine eigen-frequencies are measured, the number of damaged areas that can be identified is lower or equal to three. A similar scheme of the chromosome can be found in [43] with reference to stiffness identification problems.

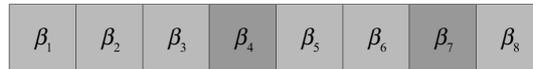

*Figure 5 - The generic chromosome*

The task of the genetic algorithm is that of exploring the space of all the possible chromosomes, in search of the damage scenario which maximizes an opportune *fitness function*, related to the objective functions and defined as follows:

$$F(C_i) = K - O_b(C_i) - g(\Sigma, N_p) \qquad (b=1,2) \tag{8}$$

where:

- the objective functions $O_1$ and $O_2$ have been defined in equations (7);

- $K$ is an arbitrary constant, chosen great enough to have $F > 0$ for every possible chromosome;

- $g(\Sigma, N_p)$ is a penalty function defined in the following.

The present study aims at identifying multiple portions of damage in the arch, with arbitrary extensions and intensities. For a given number of damaged portions $\Sigma$ the arch can be subdivided in a maximum number of parts equal to $2\Sigma+1$ depending on whether the damaged zone is located in an internal part or at the boundary of the arch. Examples of possible damage distribution of an arch with a single damage are reported in Figure 6.

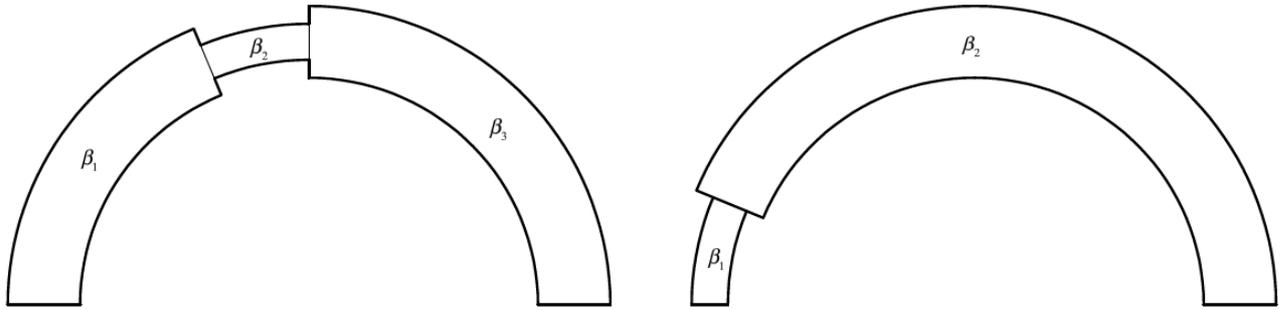

Figure 6 - The reduced arch with one damage is divided into 3 or 2 parts.

The cost function $g(\Sigma, N_p)$ allows guiding the selection of chromosomes towards the sub-space compatible with $N_p$ and is defined as follows

$$g(\Sigma, N_p) = \begin{cases} W_g\left[N_p - (2\Sigma+1)\right] & \text{if} \quad 2\Sigma+1 < N_p \\ 0 & \text{if} \quad 2\Sigma+1 \geq N_p \end{cases} \quad (9)$$

where $W_g$ is a weight assigned to this cost function. Notice that the presence in Eq. (8) of the term defined in Eq.(9) causes an increase of the fitness $F(C_i)$ not only when the value of the objective function $O_b(C_i)$ decreases, but also when the number of genes with $\beta_k > 1$ is less than or equal to the maximum expected number $N_p = 2\Sigma+1$ of arch portions. Considering the typical range of variation for both $O_b(C_i)$ and $g(\Sigma, N_p)$, in all the simulations it will be set $K = 250$ without loss of generality. Therefore, 250 will be also the highest possible value for the fitness $F(C_i)$, corresponding to the identification of the exact damage scenario.

In Figure 7 the procedure previously described is further clarified. Each chromosome $C_i$, corresponding to a generic distribution of damage along the arch, is associated to an equivalent arch where the homogeneous segments are conveniently merged aiming at an optimization of the problem size, thus making the computation of the eigen-properties of the arch faster and more effective. Then the fitness function associated to the chromosome $C_i$ can be obtained through Eq. (8). Finally, the performance of the relevant damage configuration can be assessed according to the optimization procedure described in the following.

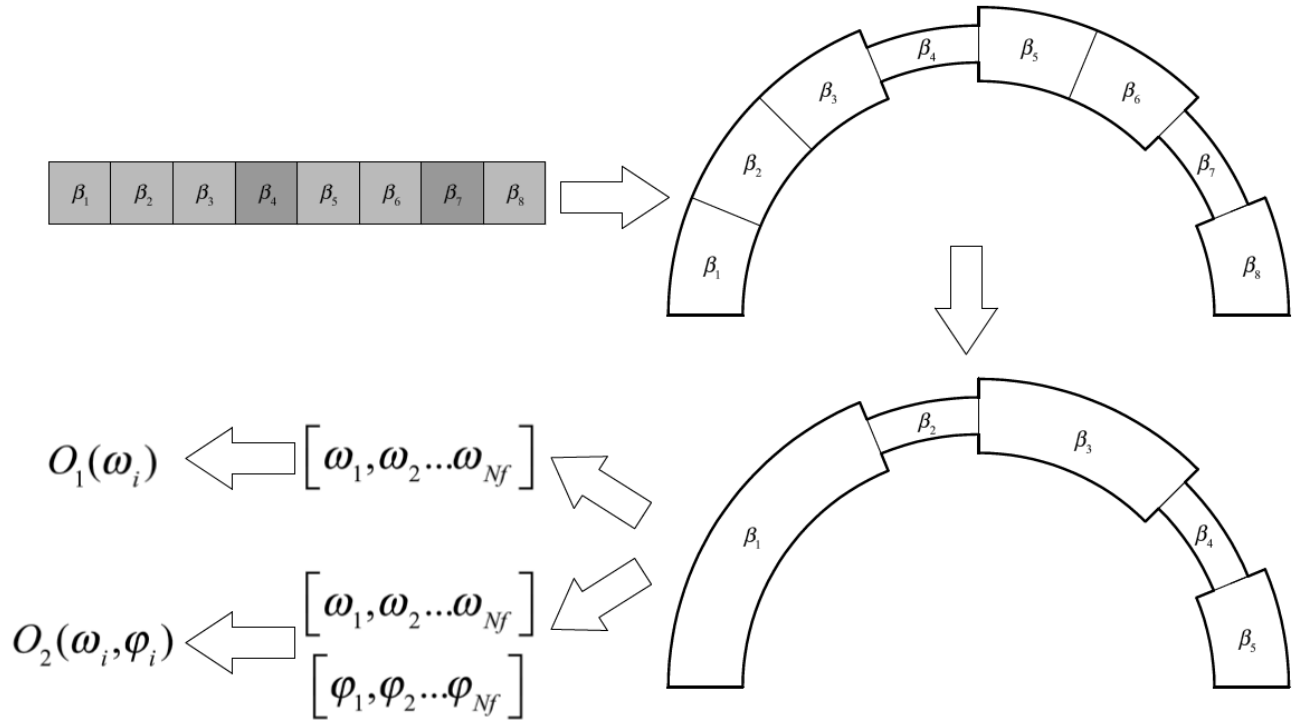

*Figure 7 - Scheme of the stages to obtain the objective functions*

The fitness value associated to each chromosome evaluates how much the corresponding damage configuration matches with the real one and represents the probability of survival of that chromosome under the pressure of the natural selection process.

More in detail, starting from the initial population of *P* chromosomes, randomly chosen among the $P_{max}$, a new generation of the same size is created from the old one, where chromosomes that have a higher fitness score are more likely to be chosen as *parents* than those that have low fitness scores. In particular in the adopted selection process, the *crossover-rate* controls the percentage of each new generation that has to be created through sexual reproduction (recombination or crossing-over between the genes of two parents' chromosomes) and the percentage (*100 – crossover-rate*) to be created through asexual reproduction (cloning of one parent's chromosome), while the *mutation-rate* controls the percentage probability of mutation, which applies to each genes of all the chromosomes in the new generation.

By iteratively repeating the selection process several times, chromosomes with the highest fitness will be progressively selected in the space of all the possible combinations and will quickly spread among the population reducing the diversity of the individuals, until (almost) only one of them will survive: hopefully, that one with the maximum fitness, that should correspond to the damage scenario more in agreement with the experimental test. Of course, it is frequent for the dynamics to remain trapped into local maxima of fitness therefore it is convenient to run the genetic algorithm many times (events), each time starting from a different initial population, in order to gain more chances to reach the global maximum of fitness. In fact, the random mutation of genes in chromosomes helps to avoid getting stuck at local optima; nevertheless, the adoption of this strategy does not prevent from getting a chromosome at the end of the run which is not actually the best one; to this regard it is worth to note that the final output of the algorithm strongly depends on the initial population randomly chosen. To increase the chance of getting the best possible chromosome, the best practice is to run more and more times the algorithm, starting every time from different initial populations.

# 6  Numerical applications

In the present section some examples of damaged arches with different boundary conditions and damage parameters have been considered. In all the numerical applications the considered damage model implies a stiffness reduction and a loss of mass. The first application refers to a hinged-hinged arch with a damage zone of considerable extent to which the proposed procedure is employed with the aim of finding the exact given damage scenario. The influence of the number of natural frequencies adopted as well as the employment of the mode shapes in the identification procedure are assessed. In particular, the two objective functions $O_1$ and $O_2$ defined in Eq. (7) are considered taking into account different numbers of eigen-properties. The performed applications allow to select the number and the type of required vibrational data which give the best performance in the identification procedure and will be therefore used in the following applications.

The setup of the simulation parameters for all the applications will always be the same one sought as the best after several applications : the size of the initial population of chromosomes, randomly chosen among the $P_{max}$, is fixed at $P = 100$ individuals; the crossover rate at 80% and the mutation rate at 2%; finally, the number of generations has been set to 100, enough to reach a stationary state for the average fitness of the chromosomes. In addition the optimal values of the weights $W_f$, $W_m$ and $W_g$ to be employed are inferred from parametric analyses reported in the following.

On the same example the influence of noise is duly assessed in the second sub-section, thus checking the robustness of the proposed approach in presence of errors on the measured data.

Then, a further example deals with a clamped-clamped arch in which the extension of the damage is small. Differently from the previous example regarding the hinged-hinged arch with extended damage, in this case the space of chromosomes does not contain the actual damage scenario and the robustness of the proposed procedure with respect to the adopted discretization is assessed. An application which takes into account the results reported in the literature with reference to an experimental test on a damaged arch is simulated in the fourth sub-section.

Finally, an example with two damaged portions is reported to demonstrate the suitability of the proposed procedure also in presence of multiple damaged portions.

In order to assess the accuracy of the results of the proposed identification procedure, three error parameters respectively of the damage location $\varepsilon_\zeta$, extension $\varepsilon_\phi$ and intensity $\varepsilon_\beta$ are introduced as follows:

$$\varepsilon_\zeta = \frac{\left|\zeta_d - \bar{\zeta}_d\right|}{\Phi}$$

$$\varepsilon_\phi = \frac{\left|\phi_d - \bar{\phi}_d\right|}{\Phi} \qquad (10)$$

$$\varepsilon_\beta = \frac{\left|\beta - \bar{\beta}\right|}{\beta}$$

being $\bar{\zeta}_d$, $\bar{\phi}_d$ and $\bar{\beta}$ the damage location, extension and intensity associated to the chromosome with the highest fitness, and $\Phi$ is the opening angle of the arch.

## 6.1 Hinged-hinged arch with an extended damaged zone

The considered arch has an opening angle $\Phi=\pi$, radius $R=1$m and circular cross section with structural parameters defined in Eqs. (2) and (4) equal to: $\lambda_1=\lambda_2=500$, $\eta=2$, $\bar{\nu}=2(1+\nu)/0.89$ $\mu=1/(1+\nu)$, . The mechanical properties are defined by means of the Young's modulus $E=206000$ MPa; Poisson's ratio $\nu=0.3$ and density $\rho=7850$ kg/m$^3$. The boundary conditions are assumed to be of simply support in the plane of the arch while clamped in the orthogonal direction.

The damage is located at the polar coordinate $0.400\,\pi$ from the left end of the arch and its intensity is $\beta=1.2$ while the extension is equal to $0.067\,\pi$. The first six frequencies of vibration and the correspondent modal shapes of the considered arch are plotted in Figure 8.

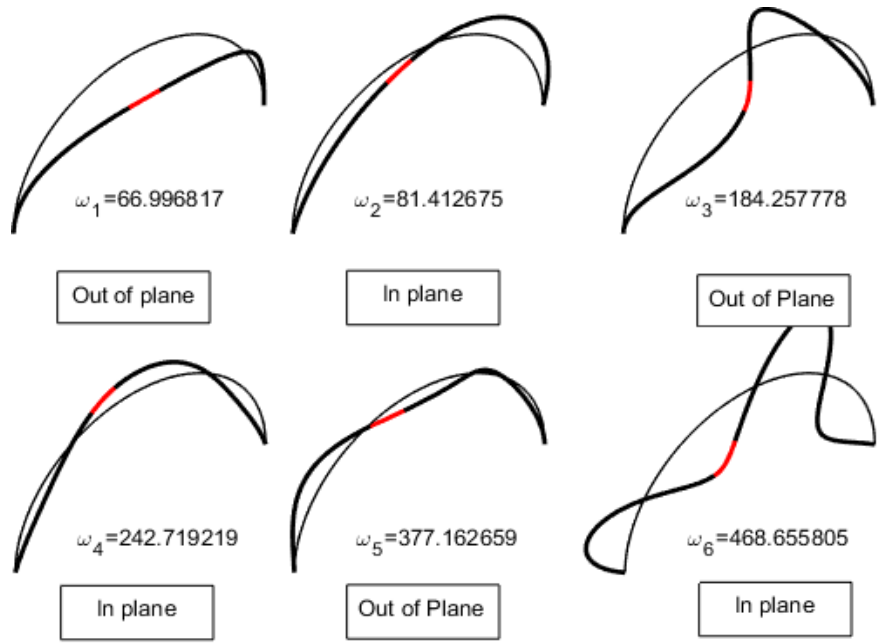

*Figure 8 -  First six modes of vibration and correspondent frequencies of the considered damaged arch*

The arch length has been divided into 30 elements; the inverse problem regarding the damage identification deals, in this case, with chromosomes with 30 genes which can assume 6 values (different intensities of damage) $\beta=[1.0,1.1,1.2,1.3,1.4,1.5]$. According to the adopted chromosome scheme, a chromosome with all genes equal to one except for the thirteenth and the fourteenth which are equal to 1.2 is able to exactly reproduce the given damage scenario.

The proposed procedure has been applied to the arch described above by considering the objective function $O_1$, that is considering only the frequency measurements, and accounting for three natural frequencies. The considered weights of the components of the fitness function are set equal to $W_f = 1$, $W_m = 0$, $W_g = 1$ and the corresponding results are shown in Figure 9.

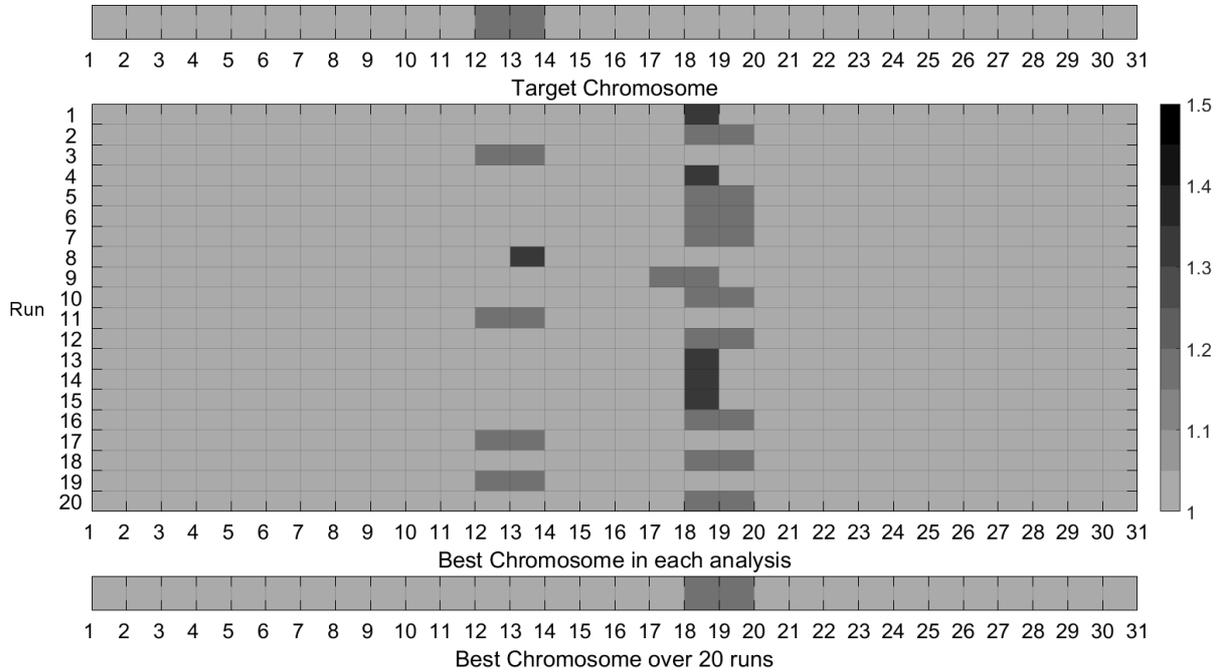

*Figure 9- Process of natural selection of the chromosome with best fitness for the objective function $O_1$ calculated with 3 natural frequencies*

The upper string represents the target chromosome which corresponds to the actual damage scenario while in the string at the bottom the chromosome with the highest fitness over 20 runs is reported. The strings in the middle represent the best chromosomes for each of the twenty runs of the genetic algorithm. The intensity of the damage is indicated with reference to the scale of grey reported in the right part of the figure.

In the objective function $O_1$ the damage parameters related to the best chromosome are: position $\zeta_d$ =0.600 $\pi$, intensity $\beta$=1.2, extension $\phi_d$=0.067 $\pi$. As it can be noticed the application leads to a best chromosome in which the damage has the correct extension and intensity but may be located at a symmetric location with respect to the actual one. This result is due to the fact that symmetric arches have the same natural frequencies when the damage is located symmetrically with respect to the vertical axis [44].

In Figure 10a the trend of the mean value of the fitness function versus the considered iteration is reported, showing how quickly the fitness function approaches the maximum value. In Figure 10b the diversity function of the population at each iteration is reported, showing that the population tends to rapidly align towards the optimal individual. In particular, to define the diversity of the population the following Hamming distance between the chromosomes will be employed [50]:

$$H_{jk} = \sum_{i=1}^{N_p} \left| \beta_i^j - \beta_i^k \right| \quad (11)$$

where $\beta_i^j, \beta_i^k$ are the $i$-th genes of the $j$-th and $k$-th chromosomes of the population, respectively. In addition, to assess how the initial population tends to evolve towards the winning chromosome, the diversity index of the population is introduced as follows

$$Diversity = \frac{2}{P(P-1)N_p(\beta_{max}-1)} \sum_{k=1}^{k=P-1} \sum_{j=k+1}^{j=P} H_{jk} \quad (12)$$

that is a parameter accounting for the mutual Hamming distance of the chromosomes belonging to a certain generation, normalized by the maximum possible diversity.

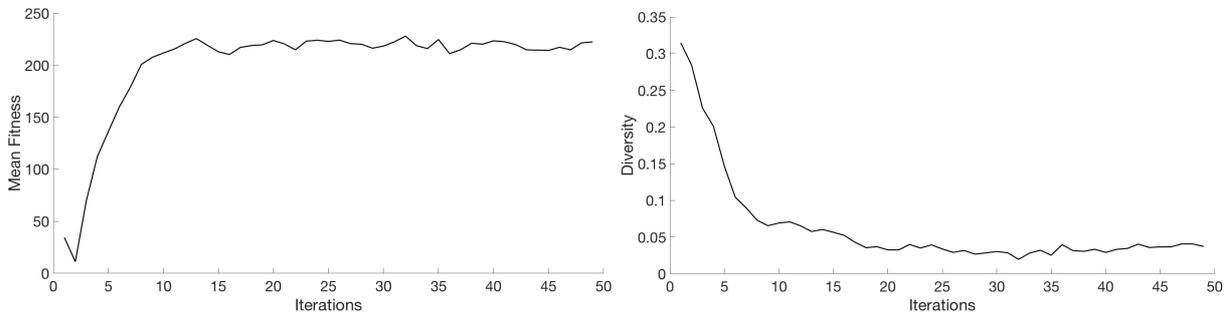

*Figure 10 - Iterations of a run vs (a) mean fitness function of the population*

*(b) diversity of the population*

In order to overcome the inconvenience on the correct location of the damage in the following Figure 11 the objective function $O_2$ has been considered in the evaluation of the fitness, and therefore both the frequencies and the modes of vibration have been taken into account, in this case the following weights have been assumed $W_f = 1$, $W_m = 1$, $W_g = 1$. The modes are evaluated in 20 equally spaced points taking into account only displacements, and avoiding the rotations which are more difficult to be measured in practical applications.

As it can be observed, in this case the correct damage position, intensity and extension has been identified as best result over 20 runs, without any symmetry ambiguity.

In the objective function $O_2$ the damage parameters related to the best chromosome are: position $\zeta_d$ =0.400 $\pi$, intensity $\beta$=1.2, extension $\phi_d$ =0.067 $\pi$; therefore for the present application the identified parameters turn out to be exact.

From the previous considerations follows that in the next applications reference will be always made to the objective function $O_2$. Although the three eigen-properties allowed recovering the correct solutions, in order to guarantee the robustness of the procedure also in presence of possible instrumental errors and in case of more than one damage, six frequencies and six modes of vibration will always be taken into account in the following applications.

In order to investigate the role of the weights $W_f$, $W_m$ and $W_g$ employed in the objective function reported in Eq. (7) a further parametric analysis is here performed. In particular, starting from the default set of parameters ($W_f$=1, $W_m$=1 and $W_g$=1), which have turned out to be reliable, each of them has been independently ranged from 0 to 5. By performing a single run for each of these sets of weights on the benchmark here proposed and considering a set of three eigenproperties as data of the algorithm, the results are reported in terms of Hamming distance with respect to the target chromosome, showing the weights do not influence that much the quality of the results unless they get far from the unit value, as shown in Figure 12.

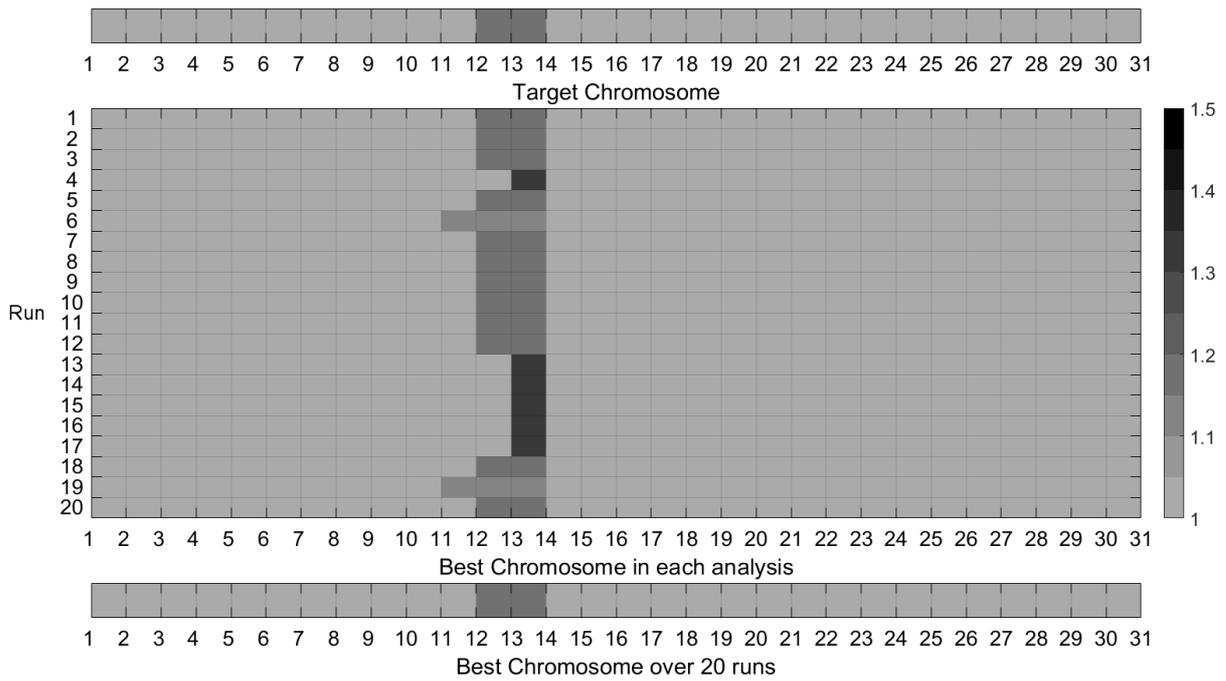

*Figure 11 - Process of natural selection of the chromosome with best fitness for the objective function $O_2$ calculated with 3 natural frequencies and 3 modes of vibration*

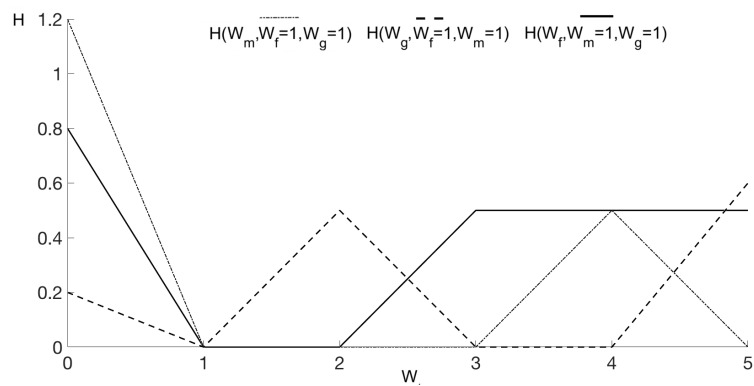

*Figure 12 - Influence of the weights adopted in the objective function on the Hamming's distance of the winning chromosome from the actual damage configuration: (a) $W_f$, (b) $W_m$ and (c) $W_g$*

### 6.2 Sensitivity of the proposed procedure to noise

A real damage identification procedure is based on the acquisition of experimental data which are affected by unavoidable noise. The purpose of the following application is to assess the performance of the proposed identification procedure when the eigen-properties of the considered damaged hinged-hinged arch are affected by errors modelled as random variables.

A simple type of random error is used to model measurement noise, i.e. proportional errors which are representative of actual measurement errors when all the instruments are set to optimal sensitivity and range, as already extensively studied with reference to cracked straight beams [51]. The natural frequencies and the modes of vibration measurements considered to perform the damage identification procedure are as follows:

$$\omega_i = \omega_i^{ex}\left(1+\varepsilon_p R\right) \tag{13}$$

$$\varphi_{i,k} = \varphi_{i,k}^{ex}\left(1+\varepsilon_p R\right) \tag{14}$$

where $\omega_i^{ex}$, $\varphi_{i,k}^{ex}$ are the exact values for the *i*-th natural frequency and the *k*-th component of the correspondent mode of vibration; $R$ is a uniformly distributed random variable in $[-1,1]$ with zero mean; $\varepsilon_p$ is a parameter governing the noise level proportional to the actual eigen-properties. Figure 13 shows the performance of the proposed procedure in damage identification when both the 6 measured frequencies and all the displacement components of the correspondent modes of vibration are affected by an increasing proportional error in the range $\varepsilon_p = 0 - 0.03$. The studied arch and the relevant damage scenario are consistent with the example treated in subsection 6.1 apart from the number of segments which is here 180. The results are reported in terms of relative error for each of the three identified parameters of the best chromosome after 20 runs in Figure 13a, and in terms of Hamming distance between the best chromosome and the target one normalized by the maximum possible Hamming distance $H_{max}=N_p(\beta_{max}-1)$ in Figure 13b. Figure 13a shows that the proposed procedure identifies with very good precision position and extension of the damage, since the error in the identified values is very similar to that assumed for the measures. Higher errors are encountered for the identified damage intensity, but the latter aspect is mainly due to the rough approximation adopted in terms of damage scale (only six levels of damage are considered). The normalized Hamming distance reported in Figure 13b represents a unified indicator of the quality of the obtained solution; again the error in the identified solution is comparable with that considered in the measures.

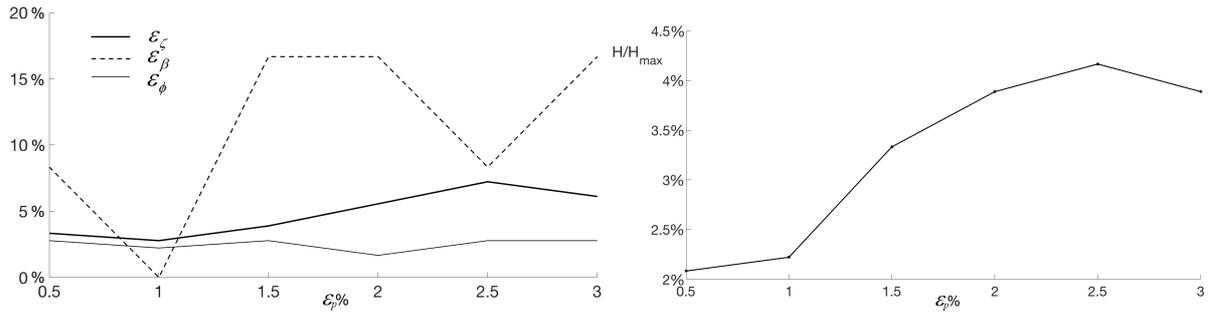

*Figure 13 - Influence of noise: (a) errors in the identified parameters and (b) normalized Hamming distance of the best solution from the target chromosome*

### 6.3 Clamped-clamped arch with a short damaged zone

The performance of the proposed damage identification procedure has been tested in this section with reference to an arch, whose ends are clamped in every direction, with opening angle $\Phi=\pi$, radius equal to 1 m and circular cross section. The structural parameters defined in Eq. (2) and (4) are equal to: $\lambda_1=\lambda_2=500$, $\eta=2$, $\bar{v}=2(1+v)/0.89$, $\mu=1/(1+v)$. The mechanical properties are defined by means of the Young's modulus $E = 206000$ Mpa; Poisson's ratio $v=0.3$ and density $\rho=7850$ kg/m$^3$. The damage is located at the polar coordinate 0.889 $\pi$ from the left end of the arch and its intensity is $\beta=1.28$ while the extension is equal to 0.016 $\pi$. The first six frequencies of vibration and the correspondent modal shapes of the considered arch are plotted in Figure 14.

In order to test the proposed procedure when no chromosome is able to accurately reproduce the actual damage scenario, in the identification process many applications dividing the arch into different number of elements have been performed. The results of the damage identification procedure, taking into account all the 6 eigen-properties, and different number of elements are reported in Figure 15.

The arch length with the indication, in dark grey, of the damaged zone is reported in the upper part of the figure. The lines reported below show the identified damage scenarios when the arch is divided into 60, 100 and 200 elements.

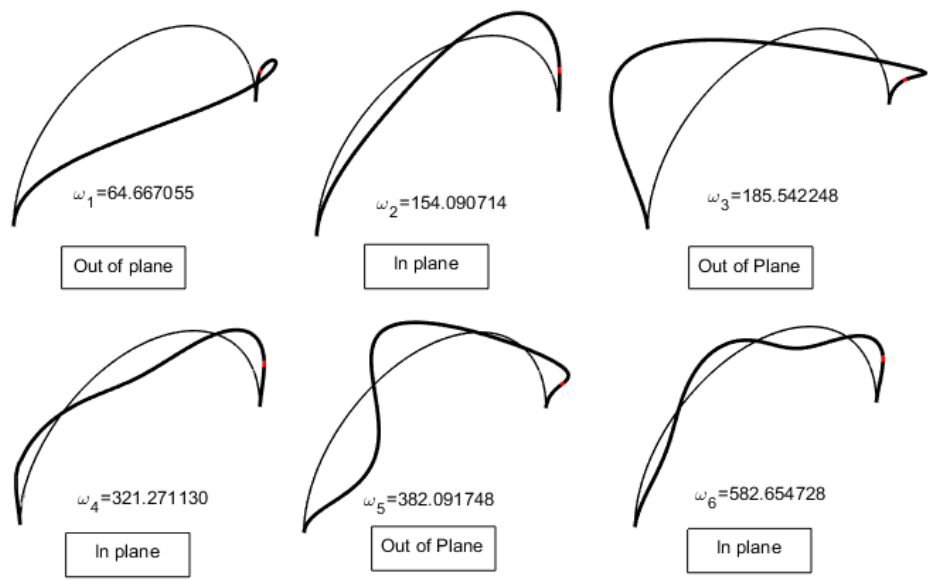

*Figure 14 - First six modes of vibration and correspondent frequencies of the considered damaged arch*

For each discretization the extension of the identified damage corresponds to the number of elements coloured in dark grey while the intensity is represented by the value of the parameter $\beta$ reported in the right part of the figure.

The best overall results are related to the discretization in 200 elements since in this case the errors on the position, the extension of the damage zone and on the intensity are respectively equal to 0.92%, 0.13% and to 1.56%; however all the results are summarized in Table 1.

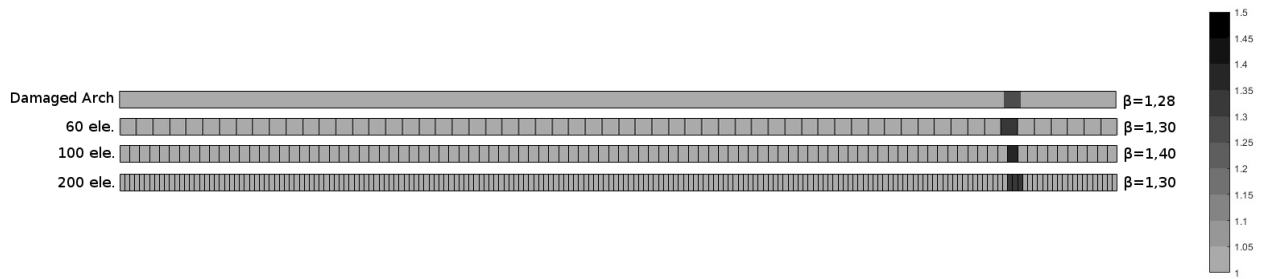

*Figure 15 - Damage identification procedure with different discretization of the arch length.*

*Table 1 - Best damage scenarios obtained with the proposed procedure according to the adopted discretization*

|  | Real damage | Adopted discretization (nr. of elements) | | | | | |
|---|---|---|---|---|---|---|---|
|  |  | 60 | | 100 | | 200 | |
|  |  | Value | Error | Value | Error | Value | Error |
| Location | $0.889\,\pi$ | $0.900\,\pi$ | 1.67% | $0.900\,\pi$ | 1.67% | $0.895\,\pi$ | 0.92% |
| Extension | $0.016\,\pi$ | $0.017\,\pi$ | 0.12% | $0.010\,\pi$ | 0.88% | $0.015\,\pi$ | 0.13% |
| $\beta$ | 1.28 | 1.30 | 1.56% | 1.40 | 9.37% | 1.30 | 1.56% |

### 6.4 Performance in the experimental case proposed by Cerri et al [52]

The proposed procedure has been successfully adopted for the identification of damage in arches characterized by both large and small extents. A further validation is here presented with reference to the experimental results obtained by Cerri et al. [52] on a steel arch affected by a concentrated damage induced by a notch. The arch is characterized by a radius equal to 1000 mm and by an angular span equal to $0.667\,\pi$, the cross section is rectangular with a base equal to 45 mm and a height equal to 15 mm. The Young's and the shear moduli are respectively equal to 206000 MPa and 80000 MPa and the density is equal to 7850 kg/m3; the notch has a depth equal to 7.5 mm and is located at the polar coordinate 76,5° from the left of the arch. All the data relative to the tested arch are reported in Table 2; the results of the experimental dynamic characterization of the arch are reported in [52]. The experimental mode shapes have been obtained through a reconstruction of the data graphically reported in [52], considering the vertical and horizontal components (no rotation) for 20 equally spaced points. The characteristics of the damage are here identified by means of the proposed procedure.

The depth of the notch can be associated to a reduction of the cross section correspondent, in the present study, to a value $\beta=2.00$. In the reported application the arch was subdivided into 1000 elements in order to guarantee the possibility for the proposed procedure to grasp the presence of a concentrated damage. The model was calibrated with the undamaged condition first, see Table 3, considering the first four frequencies. Then the parameters related to the damage were set; position and depth of the damage were given data from the experimental test, but the extension of the beam had to be calibrated.

Table 2 - Best damage scenarios obtained with the proposed procedure according to the adopted discretization

| Properties | Values |
|---|---|
| Young's Modulus (of longitudinal elasticity), $E$ | 206000 Mpa |
| Shear modulus (of tangential elasticity), $G$ | 80000 Mpa |
| Mass density, $\rho$ | 7850 kg/m$^3$ |
| Radius of the axis of the arch, $R$ | 1000 mm |
| Angular span of the arch, $2\theta$ | 120° |
| Base of the cross-section, $B$ | 45 mm |
| Height of the cross-section, $H$ | 15 mm |
| Area of the cross-section, $A$ | 675 mm$^2$ |
| Relevant moment of inertia of the cross-section, $I$ | 12656 mm$^4$ |
| Damage depth | 7.5 mm |
| Damage location | 76.5° |

The numerical model employed in [52] is in fact based on the concept of the equivalent rotational spring model, that is concentrated damage; on the other hand, the analytical model here employed [47] is based on a model of extended damage. Since in the present study the location of the damage is defined as the curvilinear abscissa of the left reduced cross section (and not its middle) with respect to the left end of the arch, the polar coordinate $0.4250\pi - 0.0005\pi = 0.4245\pi$ will be considered in the following.

The adopted mesh size, to be related to the correct extension of the damage, was therefore calibrated on the first fundamental frequency of the damaged arch $\omega_{d1}$, considering a single damage located at the section $0.4245\pi$, with intensity equal to $\beta=2.00$ and variable extension, as shown in Figure 16. The damage extension which crosses the experimental fundamental frequency is 0.003 rad that is about $0.001\pi$.

It is worth to note that, for the damaged scenario, the errors of the employed analytical procedure with respect to the experimental measures range in the interval 0.00-2.12 %, which are compatible with the noise explored in the previous sub-sections.

*Table 3 - Comparison between the experimental [52] and the analytical results obtained by means of the procedure [47] here employed, for the undamaged and damaged arches*

| Mode | Undamaged | | | Damaged | | |
|---|---|---|---|---|---|---|
| | Experimental (Hz) M.N. Cerri et al. [52] | Analytical (Hz) Caliò, Greco, D'Urso [47] | Error [%] | Experimental (Hz) M.N. Cerri et al. [52] | Analytical (Hz) Caliò, Greco, D'Urso [47] | Error [%] |
| 1 | 24.52 | 24.44 | 0.326 | 24.32 | 24.32 | 0.00 |
| 2 | 61.78 | 61.69 | 0.146 | 61.60 | 61.65 | 0.08 |
| 3 | 118.05 | 119.05 | 0.847 | 116.43 | 118.01 | 1.36 |
| 4 | 184.11 | 188.15 | 2.194 | 183.81 | 187.71 | 2.12 |

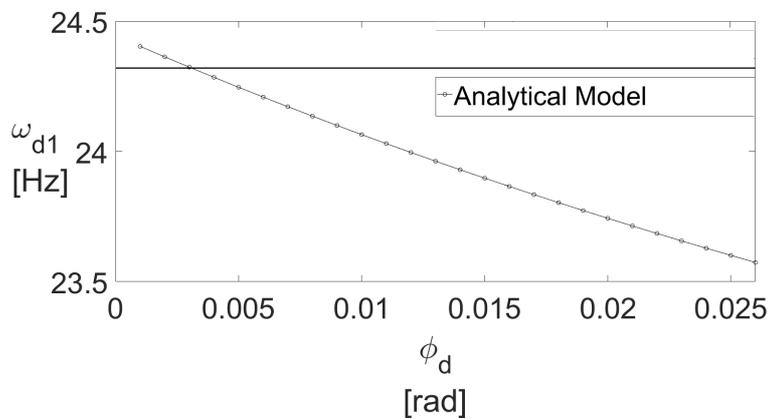

*Figure 16 - First fundamental frequency vs damage extension*

Five levels of damage have been considered, namely $\beta$=[1.0,1.5,2.0,2.5,3.0]. For the application of the proposed procedure both frequencies and modes of vibration have been adopted (considering just in-plane modes). The absolute best chromosome, over 40 performed runs, is characterized by the damage location 0.443 $\pi$, a damage extension equal to 0.002 $\pi$ and the intensity parameter $\beta = 2.00$. Such results, in terms of relative error, correspond to $\varepsilon_\zeta = 4.3\%$, $\varepsilon_\phi = 0.14\%$ and $\varepsilon_\beta = 0\%$, respectively. The obtained results demonstrate the reliability of the model also with reference to a concentrated damage and when real experimental data are employed for the identification procedure. In Figure 17 both the real and the identified damage configurations are reported, being the real damage the unsymmetrical cut and the identified one the symmetric notch. In the same Figure 17 the light grey area represents the envelope of the possible locations obtained as the best chromosomes for each of the performed 40 runs. Such envelope contains the actual damage location, demonstrating again the reliability of the proposed identification strategy. The left and right ends of the light grey portion are associated to the curvilinear abscissae 0.417 $\pi$ and 0.455 $\pi$ respectively, and correspond to an extent of the possible damaged zone $\Omega$= 0.038 $\pi$. A possible cause of differences between experimental tests and the proposed approach might be due to the fact that in the experimentally tested arch the damage was induced by an unsymmetrical notch, while the proposed approach is based on the assumption of homothetic cross section reduction, as better shown in Figure 17.

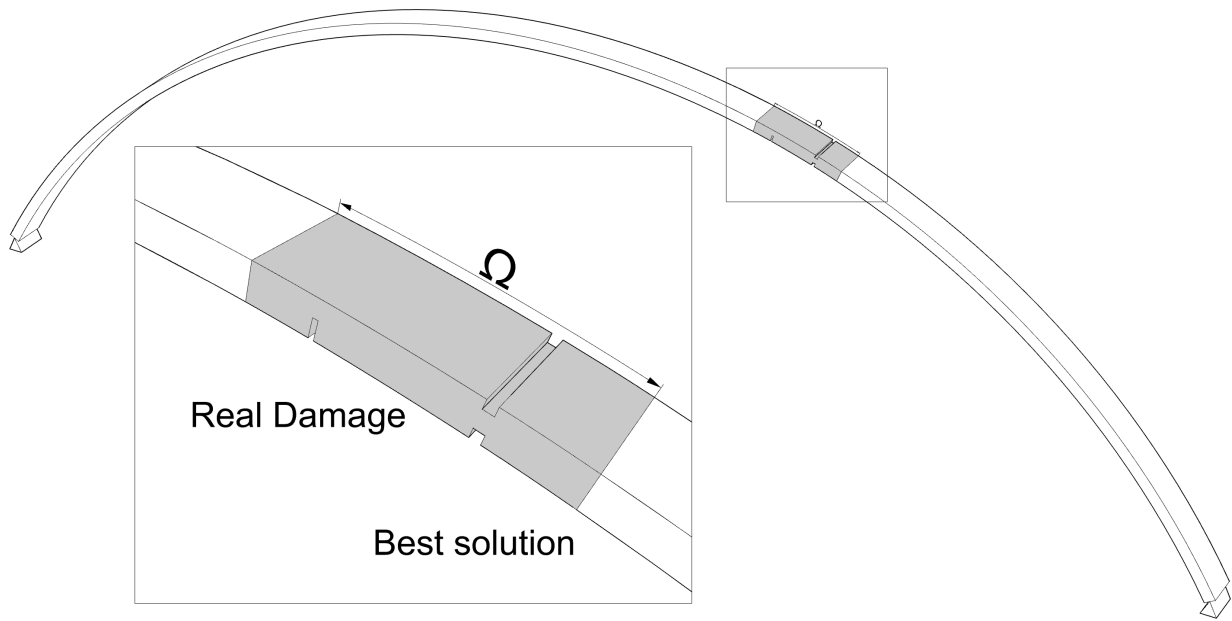

*Figure 17 - Damage identification procedure relative to an experimental case.*

**6.5   Multiple damage identification**

The last application here reported is relative to an arch with two damaged portions to demonstrate the suitability of the proposed approach also in presence of multiple damaged areas. The considered arch is characterized by ends clamped in every direction, radius equal to 1 m, opening angle $\Phi=\pi$, and circular cross section. The structural parameters defined in equations (2) and (4) are equal to: $\lambda_1=\lambda_2=500$, $\eta=2$, $\bar{v} = 2(1+v)/0.89$, $\mu=1/(1+v)$. The mechanical properties are defined by means of the Young's modulus $E = 206000$ Mpa; Poisson's ratio $v=0.3$ and density $\rho=7850$ kg/m$^3$.

Two damaged portions are considered, a shallow notch $\beta=1.10$ with a large extent 0.030 $\pi$ located at the polar coordinate 0.200 π from the left end of the arch, and a second more severe damage β=1.40 with a smaller extent 0.010 π located at the polar coordinate 0.550 π.

The first six frequencies of vibration and the correspondent modal shapes of the considered arch are plotted in Figure 18.

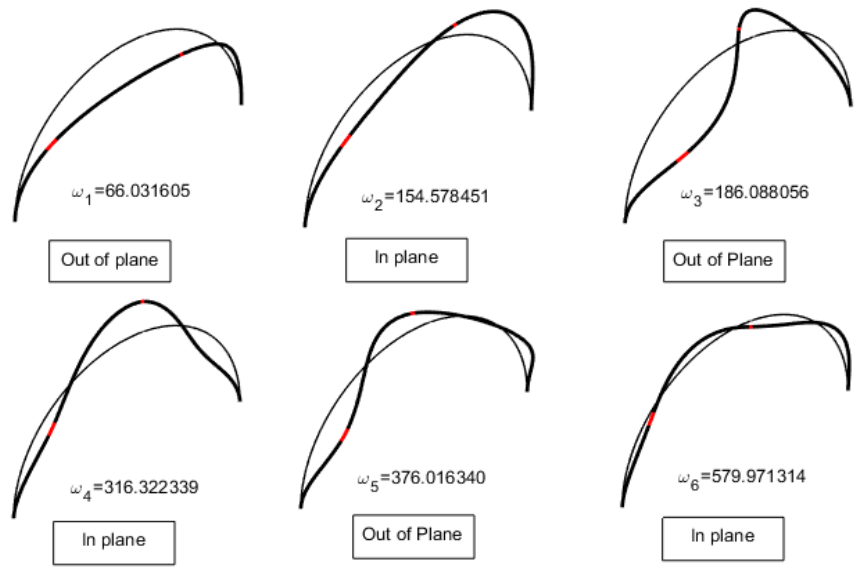

*Figure 18 - First six modes of vibration and correspondent frequencies of the considered damaged arch*

For the identification procedure the arch has been subdivided into 100 elements and 6 values of damage intensities β=[1.0,1.1,1.2,1.3,1.4,1.5], have been considered. Again, both natural frequencies and modes are employed in the identification procedure. In Figure 19 the results obtained after 65 runs are reported. The best chromosome recovers exactly ($\varepsilon_\zeta$, $\varepsilon_\phi$ and $\varepsilon_\beta$ are zero for both damage) the damage scenario to be identified, thus demonstrating also in the case of multiple damage portions the reliability of the proposed strategy. It is worth to note that a long shallow notch and a deep short one could provide similar effects on the overall behaviour of the arch; for the latter reason a larger amount of runs was needed to recover the correct damage scenario and the dispersion of the results is larger than that encountered for single damage configurations.

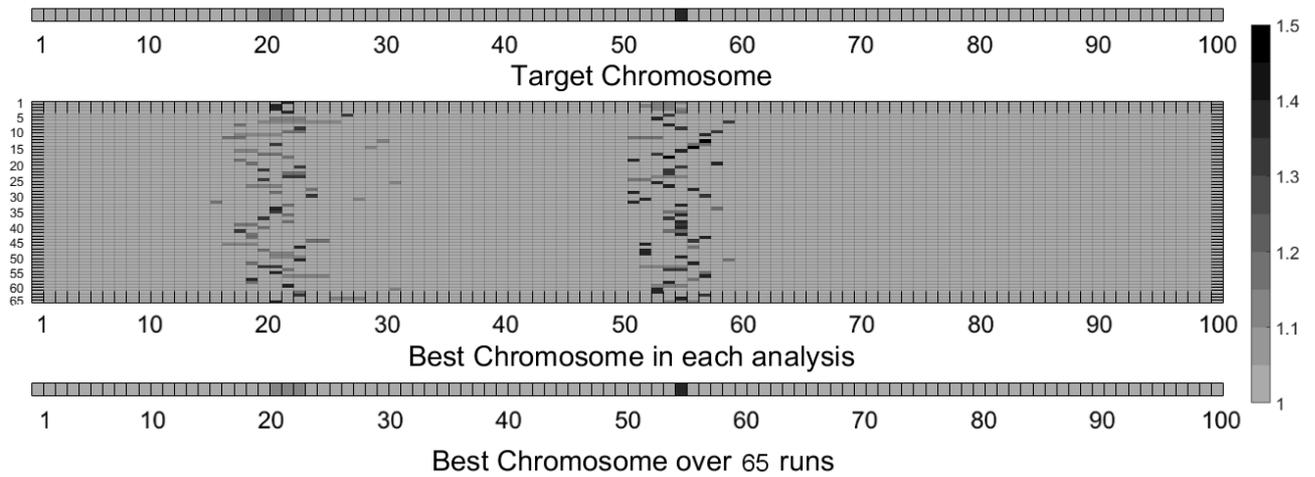

*Figure 19 - Process of natural selection of the chromosome with best fitness for the objective function $O_2$ calculated with 6 natural frequencies and 6 modes of vibration*

# 7 Conclusions

In this paper a new procedure for the damage identification in arches using dynamic properties is presented. The damage is modelled by means of a reduction of the cross section, and is able to assess not only the change of stiffness, but also possible losses of mass in the damaged part. For the latter reason the proposed procedure is suitable for damage with both small and large extents.

Two objective functions have been proposed, namely accounting for the frequency measurements only and considering frequency and mode shapes, respectively, considering the in-plane and out of plane modes. It is shown that the addition of data provided by mode shapes, although difficult to measure, might be of great help in the damage identification procedure. The search for the optimal damage scenario which better fits with the eigen-properties measurements is performed by means of a genetic algorithm.

Several validations of the proposed procedure are presented considering small and large damage extents, and investigating the severity of the damage. In addition, it is shown that the proposed procedure provides sufficiently accurate results either in the case the measurement data are affected by an increasing noise, or in a real experimental campaign. Moreover, the procedure has been also verified in presence of multiple damaged portions.

The reliability of the proposed procedure is assessed considering the errors obtained and seems to be suitable also for practical purposes. The great strength of the proposed strategy relies in its general character and its versatility. To this purpose, the adoption of the mode shapes to avoid any symmetry ambiguity that the frequencies are not able to prevent, and the possibility of identifying a large variety of possible damage configurations in arches make the proposed strategy suitable in many scenarios.

## References

[1] R.D., Adams P. Cawley, C.J. Pye, B.J. Stone, A vibration technique for non-destructively assessing the integrity of structures, J. Mech. Eng. Sci. 20 (1978) 93–100.

[2] P. Gudmundson, Eigenfrequency changes of structures due to cracks, notches and other geometrical changes, J. Mech. Phys. Solids 30(5) (1982) 339–353.

[3] G. Hearn, R.B. Testa, Modal analysis for damage detection in structures. J. Struct. Eng. 117 (10) (1991) 3042–3063.

[4] R.Y. Liang, J. Hu, F. Choy, Theoretical study of crack-induced eigenfrequency changes on beam structures, J. Eng. Mech. (ASCE) 118 (1992) 384–396.

[5] A. Morassi, Crack-induced changes in eigenparameters of beam structures, J. Eng. Mech. (ASCE) 119 (1993) 1798–1803.

[6] Y. Narkis, Identification of crack location in vibrating simply supported beams. J. Sound Vib. 172 (1994) 549–558.

[7] F. Vestroni, D. Capecchi, Damage detection in beam structures based on frequency measurements, J. Eng. Mech. (ASCE) 126(7) (2000) 761–768.

[8] A. Morassi, Identification of Two Cracks in a Simply Supported Beam from Minimal Frequency Measurements, J. Vib. Control 7 (5) (2001) 729-739.

[9] S.S. Law, Z.R. Lu, Crack identification in beam from dynamic responses, J. Sound Vib. 285 (4-5) (2005) 967–987.

[10] A. Greco, A. Pau, Damage identification in Euler frames, Comput. Struct. 92-93 (2012) 328-36.

[11] W.X. Ren, G. De Roeck, Structural damage identification using modal data. I: Simulation verification, J. Struct. Eng. (ASCE) 128 (2002) 87-95.

[12] W.X. Ren, G. De Roeck, Structural damage identification using modal data. II: Test verification, J. Struct. Eng. (ASCE) 128 (2002) 96-104.

[13] S. Caddemi, I. Caliò, Exact reconstruction of multiple concentrated damages on beams, Acta Mech. 225 (2014)

3137-3156.

[14] M. Sanayei, O. Onipede, Damage assessment of structures using static test data, AIAA journal 29 (7) (1991) 1174-1179.

[15] M. Rezaiee-Pajand, M.S. Kazemiyan, S.A. Aftabi, Static damage identification of 3D and 2D frames, Mech. Based Des. of Struc. 42 (1) (2014) 70-96.

[16] A. Greco, A. Pau, Detection of a concentrated damage in a parabolic arch by measured static displacements, Struct. Eng. Mech. 39(6) (2011) 751-765.

[17] B.K. Raghuprasad, N. Lakshmanan, N. Gopalakrishnan, K. Sathishkumar, R. Sreekala, Damage identification of beam-like structures with contiguous and distributed damage, Struct. Control and Hlth 20(4) (2013) 496-519.

[18] D. Capecchi, J. Ciambella, A. Pau, F. Vestroni, Damage identification in a parabolic arch by means of natural frequencies, modal shapes and curvatures, Meccanica, 51 (11) (2016) 2847-2859.

[19] E. Lofrano, A. Paolone, F. Romeo, Damage identification in a parabolic arch through the combined use of modal properties and empirical mode decomposition, in: Proceedings of the International Conference on Structural Dynamics, 2014, EURODYN, 2014-January, pp. 2643-2650.

[20] F. Vestroni, Dynamic Characterization and Damage Identification, in: Dynamical Inverse Problems: Theory and Application, 2011 (eds. Gladwell, G.M.L. and Morassi, A.).

[21] G.R. Irwin, Analysis of stresses and strains near the end of a crack traversing a plate, J. Appl. Mech. 24 (1957) 361-364.

[22] G.R. Irwin, Relation of stresses near a crack to the crack extension force, 9th Congr. Appl. Mech., Brussels, 1957.

[23] P.F. Rizos, N. Aspragathos, A.D. Dimarogonas, Identification of crack location and magnitude in a cantilever beam from the vibration modes, J. Sound Vib. 138(3) (1990) 381-388.

[24] G. Gounaris, A.D. Dimarogonas, A finite element of a cracked prismatic beam for structural analysis, Comput. Struct. 28 (1988) 309-313.

[25] H. Liebowitz H, W.D. Claus Jr., Failure of notched columns, Eng. Fract. Mech.1 (1968) 379-383.

[26] H. Liebowitz, H. Vanderveldt, D.W. Harris, Carrying capacity of notched column, Int. J. Solids Struct. 3 (1967) 489-500.

[27] W.M. Ostachowicz, C. Krawczuk, Analysis of the effect of cracks on the natural frequencies of a cantilever beam, J. Sound Vib. 150(2) (1991) 191-201.

[28] S.A. Paipetis, A.D. Dimarogonas, Analytical Methods in Rotor Dynamics Elsevier Applied Science, London, 1986.

[29] P. Gudmnundson, The dynamic behaviour of slender structures with cross-sectional cracks, J. Mech. Phys. Solids 31 (1983) 329–345.

[30] J.K. Sinha, M.I. Friswell, S. Edwards, Simplified models for the location of cracks in beam structures using measured vibration data J. Sound Vib. 251(1) (2002) 13–38.

[31] L.B. Freund, G. Hermann, Dynamic fracture of a beam or plate in plane bending, J. Appl. Mech. 76 (1976) 112-116.

[32] F. Cannizzaro, A. Greco, S. Caddemi, I. Caliò, Closed form solutions of a multi-cracked circular arch under static loads, Int. J. Solids Struct. 121 (2017) 191-200.

[33] I. Caliò, A. Greco, D. D'Urso, Structural models for the evaluation of eigen-properties in damaged spatial arches: a critical appraisal, Arch. Appl. Mech. 86 (11) (2016) 1853-1867.

[34] I. Caliò, A. Greco, D. D'Urso, Damage identification on spatial arches, in: Proceedings of the International Conference on Structural Dynamic, 2014, EURODYN, 2014-January, pp. 2517-2523.

[35] Z. Nie, J. Zhao, H. Ma, L. Cheng, Visualization of damage detection for circular arch based on stochastic subspace identification, in: Proceedings - International Conference on Artificial Intelligence and Computational Intelligence, 2010, AICI 2010, 2, art. no. 5655132, pp. 127-130.

[36] A. Pau, A. Greco, F. Vestroni, Numerical and experimental detection of concentrated damage in a parabolic arch by measured frequency variations, J. Vib. Control, 17(4) (2010) 605–614.

[37] D. Dessi, G. Camerlengo, Damage identification techniques via modal curvature analysis: Overview and comparison Mech. Syst. Signal Pr. 52-53 (1) (2015) 181-205.

[38] M.I. Friswell, J.E.T. Penny, G. Lindfield, The location of damage from vibration data using genetic algorithms, 1995 IMAC XIII – 13th International Modal Analysis Conference

[39] M.T. Vakil-Baghmisheh, M. Peimani, M.H. Sadeghi, M.M. Ettefagh, Crack detection in beam-like structures using genetic algorithms, Appl. Soft Comput. 8(2) (2008) 1150-1160.

[40] L. Faravelli, F. Marazzi, Stiffness matrices and genetic algorithm identifiers toward damage detection, in: Proceedings of IABMAS'06, 2006, Porto, Portugal, (CD-ROM).

[41] M. Mehrjoo, N. Khaji, M. Ghafory-Ashtianyc, Application of genetic algorithm in crack detection of beam-like structures using a new cracked Euler–Bernoulli beam element, Appl. Soft Comput. 13 (2013) 867-880.

[42] M. Mehrjoo, N. Khaji, Crack detection in a beam with an arbitrary number of transverse cracks using genetic algorithms, J. Mech. Sci. Technol. 28(3) (2014) 823-836.

[43] S. Casciati, Stiffness identification and damage localization via differential evolution algorithms, Struct. Control Hlth 15 (2008) 436-449.

[44] I. Caliò, D. D'Urso, A. Greco The influence of damage on the eigen-properties of Timoshenko spatial arches, Comput. Struct. 190 (2017) 13-24.

[45] F.W. Williams, W.H. Wittrick, An automatic computational procedure for calculating natural frequencies of skeletal structures, Int. J. Mech. Sci. 12 (9) (1970) 781-791.

[46] W.P. Howson, A.K. Jemah, Exact dynamic stiffness method for planar natural frequencies of curved Timoshenko beams, P. I. Mech. Eng. C_J. Mec. 213 (7) (1999) 687-696.

[47] I. Caliò A. Greco, D. D'Urso, Free vibrations of spatial Timoshenko arches J. Sound Vib., 333 (19) (2014) 4543-4561.

[48] J.R. Banerjee, Dynamic stiffness formulation for structural elements: A general approach Comput. Struct. 63 (1) (1997) 101-103.

[49] R.J. Allemang, The modal assurance criterion - twenty years of use and abuse, Sound Vib., 37(8) (2003) 14-20.

[50] R.W. Morrison, K.A. De Jong, Measurement of population diversity, Lecture Notes in Computer Science (including subseries Lecture Notes in Artificial Intelligence and Lecture Notes in Bioinformatics), 2310, (2002) 31-41

[51] S. Caddemi, A. Greco, The influence of instrumental errors on the static identification of damage parameters for elastic beams, Comput. Struct. 84(26-27) (2006) 1696-1708.

[52] M.N. Cerri, M. Dilena, G.C. Ruta, Vibration and damage detection in undamaged and cracked circular arches: Experimental and analytical results, J. Sound Vib., 314 (1-2) (2008) 83-94.